\documentclass[NJP]{iopart}
\usepackage[latin9]{inputenc}
\usepackage{array}
\usepackage{amstext}
\usepackage{graphicx}
\usepackage{esint}
\usepackage[unicode=true,pdfusetitle,
 bookmarks=true,bookmarksnumbered=false,bookmarksopen=false,
 breaklinks=false,pdfborder={0 0 1},backref=false,colorlinks=false]
 {hyperref}

\makeatletter

\providecommand{\tabularnewline}{\\}

\usepackage{iopams}
\usepackage{setstack}

\makeatother

\begin{document}
\title{Pressure-driven wrinkling of soft inner-lined tubes}
\author{Benjamin Foster}
\ead{ben\_foster@berkeley.edu}
\address{Physics Department, University of California at Berkeley, Berkeley,
California 94720, USA}
\author{Nicol\'{a}s Verschueren}
\ead{nverschueren@berkeley.edu}
\address{Physics Department, University of California at Berkeley, Berkeley,
California 94720, USA}
\author{Edgar Knobloch}
\ead{knobloch@berkeley.edu}
\address{Physics Department, University of California at Berkeley, Berkeley,
California 94720, USA}
\author{Leonardo Gordillo}
\ead{leonardo.gordillo@usach.cl}
\address{Departamento de F\'{i}sica, Facultad de Ciencia, Universidad de Santiago
de Chile, Chile}
\begin{abstract}
A simple equation modelling an inextensible elastic lining of an inner-lined
tube subject to an imposed pressure difference is derived from a consideration
of the idealised elastic properties of the lining and the pressure and
soft-substrate forces. Two cases are considered in
detail, one with prominent wrinkling and a second one in which wrinkling is
absent and only buckling remains. Bifurcation diagrams are computed via
numerical continuation for both cases. Wrinkling, buckling,
folding, and mixed-mode solutions are found and organised according
to system-response measures including tension, in-plane compression,
maximum curvature and energy. Approximate wrinkle solutions are constructed
using weakly nonlinear theory, in excellent agreement with numerics.
Our approach explains how the wavelength of the wrinkles is selected
as a function of the parameters in compressed wrinkling systems and
shows how localised folds and mixed-mode states form in secondary
bifurcations from wrinkled states. Our model aims to capture the wrinkling
response of arterial endothelium to blood pressure changes but applies much
more broadly.
\end{abstract}
\noindent{\it Keywords\/}: {\noindent Nonlinear elastica; wrinkling; buckling; bifurcation}

\maketitle
\submitto{\NJP} 

\section{Introduction}

Lateral compression of a finite thin floating elastic sheet generates
periodic wrinkles whose wavelength is the result of a balance between
elastic forces and the restoring weight of the entrained liquid. On
further compression, the sheet undergoes a transition from the wrinkled
state to one characterised by a single fold \cite{pocivavsek2008}.
However, wrinkling is not exclusive to floating elastica: the weight
of the liquid can be replaced by other forces and used to generate
wrinkling in both two-dimensional circular and three-dimensional spherical
and curved geometries. Examples are provided by laterally compressed
\cite{Brau:2013jn} or curved bilayer materials \cite{stoop2015},
as well as vertically loaded floating circular sheets~\cite{roman2010,king2012,box2019}
and spring-loaded interfaces \cite{michaels_puckering_2021}. In contrast,
compressed or deflated spherical shells \cite{Vliegenthart_spherical_shell,marthelot2017reversible}
exhibit buckling with no preferred length scale, as do elastic rings
supporting a soap film \cite{katifori2009,box2020,kodio_weak_nonlin,giomi2012}.
Constrained buckling of elastic rings exhibits similar properties \cite{hazel_mullin_2017}.

Understanding how surfaces wrinkle and then fold in different geometries
under specific forces usually requires solving complicated systems
of partial differential equations. The thin floating sheet in one
dimension (1D) provides an exception. This system is not only modelled
by a simple equation for in-plane deformations, but also turns out
to be completely integrable in the limit of infinite extent
\cite{Audoly:2011cq,diamant_witten2011,rivetti2013,Oshri:2015bi,LeoEdgarsheet}.
As a result the remarkable shapes of both wrinkles and folds on thin
floating sheets can be described using stunningly simple mathematical
expressions \cite{diamant_witten2011,rivetti2013}, which naturally
implies closed formulas for the wrinkling/folding thresholds in parameter
space.

In this article we study the competition between in-plane wrinkling and buckling
in a circular geometry within a similar framework. The results lead to greater
understanding of a number of different systems where such competition is present.
These include in-plane wrinkling of the elastic lining of an artery where
wrinkled-to-unwrinkled cycles driven by diastolic-to-systolic blood pressure changes
may prevent clogging and adhesion of platelets via large changes in the
local curvature of its endothelium~\cite{pocivavsek2018,pocivavsek2019}.
Such cycling is likely to prove useful in other applications. A similar wrinkling
instability is present in a rotating Hele-Shaw cell when a higher density fluid in the center is separated from a lower density fluid on the outside by an
elastic membrane
\cite{Carrillo1996,Carvalho2014_HS_elasticA,Carvalho2014_HS_elasticB}.

We construct an idealised two-dimensional model for this class of systems and compute
strongly deformed states up to the point of self-contact, analyse their stability, and
organise the results in the form of bifurcation diagrams. These diagrams describe the
response of the system (compression, tension, maximum curvature) as a function of a
control parameter, for example, the imposed pressure difference. We use the results
to identify a transition from unwrinkled to periodic wrinkled states and then to folded
states similar to what is observed in spring-loaded linings or tubular chitosan hydrogel
surfaces \cite{michaels_puckering_2021, kumar_2021}. Fold states arise via secondary
bifurcations from the wrinkled state as in the one-dimensional case. Two cases are considered
in detail, one with prominent wrinkling and a second one in which wrinkling is absent
and only buckling remains.
\begin{figure}
\centering{}\includegraphics[width=0.75\linewidth]{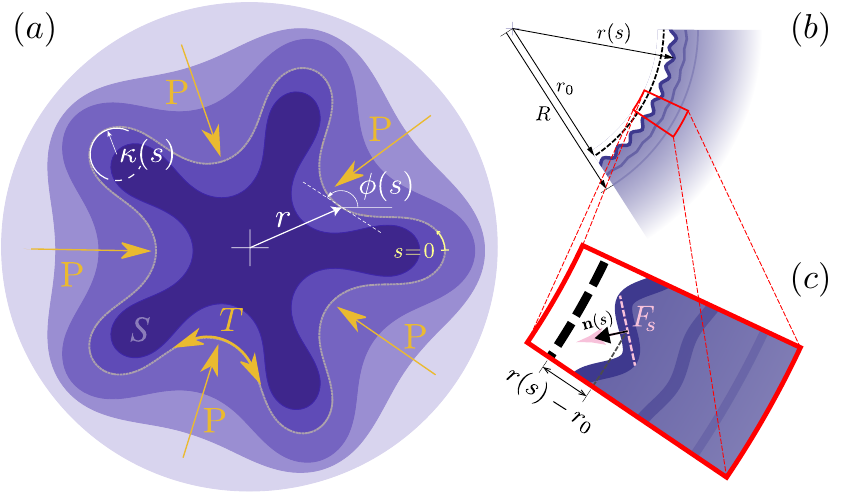}
\caption{(a) Schematic view of a tube undergoing pressure-driven wrinkling
with wavenumber $m=5$. (b,c) Force exerted on the lining by the exterior
substrate. \label{fig:model}}
\end{figure}

\section{The Model}

To represent the lining on the inside of a soft tube, we consider
an inextensible, infinitely thin membrane of length $L=2\pi R$ attached
to a soft substrate as shown in figure~\ref{fig:model}. We suppose
that in equilibrium ($P=0$) the unlined soft tube has an inner radius $r_{0}<R$
(figure~\ref{fig:model}(b)) and hence that, when lined, the lining is
forced to wrinkle. We model this force by an inward normal force per
unit area ${\bf F}_{s}=\frac{1}{2}K\left(r(s)^{2}-r_{0}^{2}\right){\bf n}\left(s\right)$
(figure~\ref{fig:model}(c)). Here $r=r(s)$ denotes the lining
profile ($r$ is the distance from the tube centre) and $s$ is the
arclength. The substrate force ${\bf F}_{s}$ is the simplest nonlinear model 
that is differentiable at $r=0$ and that behaves like the classical
Winkler foundation \cite{winkler} when expanded around $r_{0}$, with
constant stiffness $k=Kr_{0}$. Moreover, the quadratic contribution to
the force vanishes in the flat-foundation limit, i.e. as $r_{0}\rightarrow\infty$,
again recovering a Winkler-type foundation response. Although higher-order models
\cite{nguyen_2020}, and in particular models that include nonlocal contributions
\cite{Brau:2013jn}, may provide a more realistic representation of the substrate
forces, the Winkler model has been used extensively in studies of
substrate-supported elastica and has provided important insights into
the instabilities responsible for both wrinkled and localised states
\cite{michaels_puckering_2021,hunt_1989,michaels_2019}.


For in-plane deformations the resulting system is then described by
\begin{equation}
\mathcal{B}\left(\frac{1}{2}\kappa^{3}+\partial_{s}^{2}\kappa\right)-T\kappa-P+\frac{1}{2}K\left(r_{0}^{2}-r^{2}\right)=0,\label{eq:main}
\end{equation}
where $\kappa\equiv\partial_{s}\phi$ is the local curvature. Here $\phi$
is the angle between the tangent plane and the horizontal or $x$-axis (figure \ref{fig:model}(a)).
In terms of Cartesian coordinates $(x,y)$ with origin at the tube center,
$\partial_{s}x=\cos\phi$, $\partial_{s}y=\sin\phi$ and $r^{2}\equiv x^{2}+y^{2}$.
The constants in (\ref{eq:main}) are the bending modulus $\mathcal{B}$ and
the (unknown) tension $T$ required to maintain the length $L$ of the
lining ($T<0$ implies tangential compression). A brief derivation
of (\ref{eq:main}) similar to that in \cite{kodio_weak_nonlin} can
be found in \ref{sec:AppendixA}. In the following we absorb the
constant term $\frac{1}{2}Kr_{0}^{2}$ in the pressure $P$. The resulting system is then similar to a rotating Hele-Shaw cell filled with two
fluids separated by an elastic membrane, with a higher density interior \cite{Carrillo1996,Carvalho2014_HS_elasticA,Carvalho2014_HS_elasticB}.

We define the natural length scale 
\begin{equation}
\lambda\equiv\left(\frac{\mathcal{B}}{K}\right)^{\frac{1}{5}},\label{eq:natural_wavelength}
\end{equation}
and introduce a dimensionless parameter that measures the perimeter
of the lining in terms of $\lambda$, $\ell\equiv R/\lambda$. We
scale (\ref{eq:main}) according to $s\sim R$, $\kappa\sim R^{-1}$,
$r\sim R$, $T\sim\mathcal{B}/R^{2}$, $P\sim\mathcal{B}/R^{3}$,
yielding 
\begin{equation}
\partial_{s}^{3}\phi+\frac{1}{2}\left(\partial_{s}\phi\right)^{3}-T\partial_{s}\phi-P-\frac{1}{2}\ell^{5}r^{2}=0.\label{eq:phiODE}
\end{equation}
The area within the lining, scaled relative to the area of the circle,
is conveniently written via Stokes theorem as 
\begin{equation}
S=\frac{1}{2\pi}\oint\left[x\sin\phi-y\cos\phi\right]\mathrm{d}s,\label{eq:area}
\end{equation}
and, accordingly, its compression is $\Delta\equiv 1-S$. The total
energy, also scaled relative to the circle, is given by 
\begin{equation}
E=\frac{2}{\pi\left(4+\ell^{5}\right)}\oint\left[\left(\partial_{s}\phi\right)^{2}+\frac{1}{4}\ell^{5}r^{2}\left(x\sin\phi-y\cos\phi\right)\right]\mathrm{d}s.\label{eq:energy}
\end{equation}

\section{Linear and weakly nonlinear theory}

The simplest solution to (\ref{eq:phiODE}) is the circle:
\begin{equation}
\phi_{0}\left(s\right)=s+\pi/2,\quad x_{0}\left(s\right)=\cos s,\quad y_{0}\left(s\right)=\sin s.\label{eq:circle}
\end{equation}
This solution requires a simple relationship between the imposed pressure and
the resulting tension, 
\begin{equation}
T_{0}=\frac{1}{2}\left(1-\ell^{5}\right)-P_{0},\label{eq:PT0}
\end{equation}
and serves as the starting point (order zero) for linear and weakly
nonlinear analysis. Introducing a small parameter $\epsilon$ measuring
the amplitude of a perturbation of the circle solution, we expand
$\phi$, $x$, $y$, $T$ and $P$ as follows: 
\begin{eqnarray*}
\phi(s) & =\sum_{j=0}^{N}\epsilon^{j}\phi_{j}(s),\quad x(s)=\sum_{j=0}^{N}\epsilon^{j}x_{j}(s),\quad y(s)=\sum_{j=0}^{N}\epsilon^{j}y_{j}(s),\\
 & T=\sum_{j=0}^{N}\epsilon^{2j}T_{2j},\quad P=\sum_{j=0}^{N}\epsilon^{2j}P_{2j}.
\end{eqnarray*}
The coefficients of odd powers of $\epsilon$ in $P$ and $T$ vanish owing to
the invariance of the system under rotations by half a wavelength. Substituting
these expansions into (\ref{eq:phiODE}) and the equations for $x$ and
$y$ leads, at $\mathcal{O}(\epsilon)$,
to 
\[
\mathcal{L}[\phi_{1},x_{1},y_{1}]\equiv\partial_{s}^{3}\phi_{1}+\left(\frac{3}{2}-T_{0}\right)\partial_{s}\phi_{1}-\ell^{5}\left(x_{1}x_{0}+y_{1}y_{0}\right)=0.
\]
To eliminate $x_{1}$ and $y_{1}$, we compute $(\partial_{s}^{2}\mathcal{L}+\mathcal{L})[\phi_{1},x_{1},y_{1}]$:
\[
\partial_{s}^{5}\phi_{1}+\left(2+P_{0}+\frac{\ell^{5}}{2}\right)\partial_{s}^{3}\phi_{1}+\left(1+P_{0}+\frac{3\ell^{5}}{2}\right)\partial_{s}\phi_{1}=0.
\]
This equation reduces to an algebraic equation for the wavenumber $m$ on assuming that $\phi_{1}(s)\propto\sin(ms+\delta)$:
\begin{equation}
m^{5}-\left(2+P_{0}+\frac{\ell^{5}}{2}\right)m^{3}+\left(1+P_{0}+\frac{3\ell^{5}}{2}\right)m=0.\label{eq:quintic_m}
\end{equation}
Modes with $m=0$ (axisymmetric expansion) and $m=1$ (translations) are excluded by inextensibility and pinning, respectively. Thus $m^\geq 2$ and solutions with integers $m$ correspond to periodic states we refer to as wrinkles (W$_m$); $\delta$ corresponds to a rigid rotation of the solution, and can be set to zero. Thus 
\begin{eqnarray}
\qquad\phi_{1} & = & \sin(ms),\label{eq:phi1}\\
x_{1}+iy_{1} & = & \frac{m\cos\left(ms\right)-i\sin\left(ms\right)}{m^{2}-1}\,\exp\left(is\right).\label{eq:phi2}
\end{eqnarray}

\begin{figure}
\centering{}\includegraphics[width=0.75\linewidth]{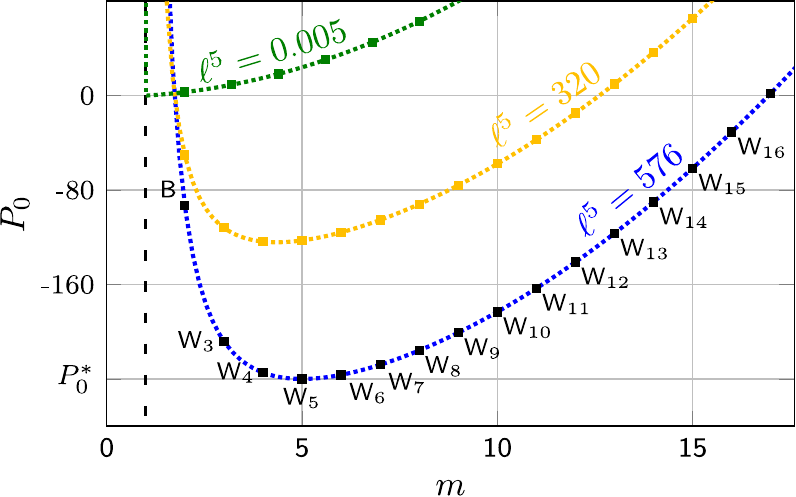}
\caption{The wrinkle wavenumber $m$ as a function of the pressure
$P_{0}$ for $\ell^{5}=576$ (blue curve, $m^{*}=5$), $\ell^{5}=320$
(yellow curve, $m^{*}=4$) and $\ell^{5}=0.005$ (green curve, $m^{*}=2$). \label{fig:P0}}
\end{figure}

Equation~(\ref{eq:quintic_m}) is an important expression as it can
be used to determine the critical pressure $P^{*}_0$ for the onset of the wrinkling
instability as the pressure increases and the wavenumber $m=m^*$ of the resulting wrinkles for a given
$\ell$. Figure~\ref{fig:P0} depicts $P_{0}$ as a function of
$m$ for three different $\ell$ values. The figure shows how the circular tube
becomes wrinkled as $P_{0}$ overcomes the threshold $P_{0}^{*}\equiv(-\ell^{5}+4\ell^{5/2})/2$
and the interior depressurises. It also shows how the choice of $\ell$
determines the order of appearance of new unstable wavenumbers. A simple
formula gives the critical wavenumber at $P_{0}^{*}$: $m^{*}=\sqrt{1+\sqrt{\ell^{5}}}$.
When $\ell^5<9$ the onset wavenumber is $m^*=2$ since $m=1$ corresponds to translations
(figure~\ref{fig:P0}).

In terms of physical parameters, 
\begin{equation}
P_{0}^{*}=\frac{1}{2}K\left(r_{0}^{2}-R^{2}\right)+2\left(\frac{\mathcal{B}K}{R}\right)^{1/2},\label{eq:pre-stretched-tube}
\end{equation}
providing a key formula relating the critical pressure $P_{0}^{*}$ for the onset of wrinkling
to the geometry of the tube and the physical properties of the substrate and the lining.
Expression (\ref{eq:pre-stretched-tube})
also indicates that the critical pressure can be tuned by a proper
choice of $r_{0}$ and $R$, for instance, to generate lining wrinkles
at pressure equilibrium ($P_{0}=0$). As mentioned, this requires
$r_{0}<R$, i.e. that the lining has an excess of length over the
unlined tube inner perimeter. Likewise, for large $\ell$, the critical
wavelength of the wrinkles in terms of physical parameters simplifies
to $\lambda^{*}=2\pi R/m^{*}=2\pi\left(\mathcal{B}/\left[KR\right]\right)^{1/4}$,
where $KR$ can be identified with the foundation stiffness $k$ if
$R\approx r_{0}$.

\begin{figure*}[t]
\centering{}\includegraphics[width=1\linewidth]{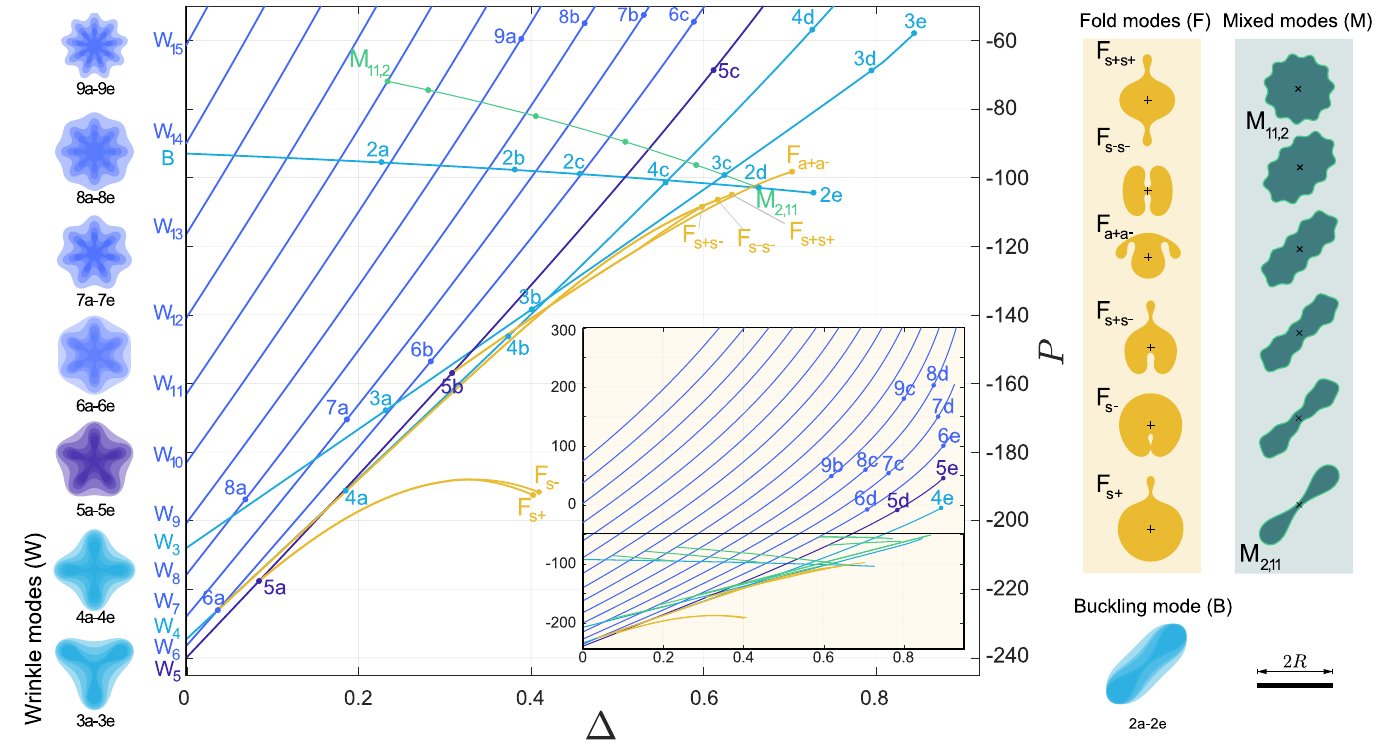}
\caption{Bifurcation diagram for $\ell^{5}=576$ (corresponding to $m^{*}=5$) showing the
  compression $\Delta$ as a function of the pressure $P$. The unperturbed circle state
  corresponds to $\Delta=0$; the primary branch W$_5$ corresponding to states with wavenumber
  $m=5$ is shown as a thin purple line. Subsequent primary wrinkle solutions W$_m$ are labelled
  by their wavenumber ($m<5$, cyan; $m>5$, blue), while the secondary solutions are labelled F or M according
  to their type (fold or mixed mode). Branches are presented up to the point of self-contact.
  Sample solutions at the locations indicated in the main plot are shown alongside with the
  different wrinkle profiles for each $m$ superposed. All solutions are reflected across the
  $x$ axis for ease of visualization (solutions F and M have been rotated by $90^{\circ}$ and
  $45^{\circ}$ for convenience; the $+/\times$ symbols at the center of each profile indicate
  the orientation of the axes). The mixed mode branches extend between M$_{m_1,m_2}$ where the
  first subscript indicates the primary wavenumber and the second the new wavenumber introduced
  at the secondary bifurcation. The subscripts $\pm$ on F refer to the folded states with an extrusion
  ($+$) or intrusion ($-$). The subscripts {\it s} and {\it a} indicate whether these protrusions occur
  on the axis or off it. The letter $\text{B}$ labels the buckling mode $m=2$. A scale bar of unit
  length is included on the right. The inset shows the same results but over a larger range of $P$.
  \label{fig:PvsDelta}}
\end{figure*}

We extend the above approach to compute periodic states with wavenumber $m$ to higher order in $\epsilon$
(see \ref{sec:AppendixB}). We display the $\mathcal{O}(\epsilon^{2})$ expressions for $\phi_{2}$,
$x_{2}$, $y_{2}$, $P_{2}$ and $T_{2}$ below: 
\begin{eqnarray*}
\qquad\phi_{2} & = & \frac{1}{8m}\sin(2ms),\\
x_{2}+iy_{2} & = & \bigg[-\frac{1}{4}+\frac{i}{8m}\sin(2ms)\biggl]\,\exp\left(is\right),\\
\qquad P_{2} & = & \frac{2m^{4}-9m^{2}+3}{8\left(m^{2}-1\right)^{2}}\ell^{5}+\frac{3\left(m^{2}-1\right)}{8},\\
\qquad T_{2} & = & \frac{3}{8\left(m^{2}-1\right)}\ell^{5}+\frac{3\left(m^{2}+1\right)}{8}.
\end{eqnarray*}
We computed the expansion to $\mathcal{O}\left(\epsilon^{7}\right)$
using computer algebra. From these results, we can compute the slope
$\partial P/\partial T$ of the primary wrinkle branches at the bifurcation
points given by (\ref{eq:quintic_m}). This slope is always positive
unless $m=2$ and $\ell^{5}\geq81$. The mode $m=2$ is special, because
of its maximum wavelength; this mode is the first one to emerge in
the absence of the intrinsic scale $\ell$ \cite{flaherty_1972},
and we therefore refer to it as the buckling mode ($\mathrm{B}$).

In the following we extend the above results using numerical continuation
and consider two cases. In the first (Section 4) substrate forces are substantial and
wrinkling is present. In the second (Section 5) these forces are much weaker, wrinkling
is absent and only buckling remains.

\section{Numerical continuation: $\ell^5=576$}

To compute strongly nonlinear solutions, we implemented (\ref{eq:phiODE}) as
a boundary value problem in AUTO \cite{doedel08auto-07p} (see \ref{sec:AppendixC}
for details) and numerically continued different wrinkle states for
a given $\ell$ starting from the circle branch satisfying (\ref{eq:PT0}).
Each increment in $P$ requires the solution of a nonlinear eigenvalue
problem for the response $T$. The results show that the weakly nonlinear
theory is remarkably accurate, even when $\epsilon=\mathcal{O}(1)$
(see \ref{sec:AppendixC} for a comparison up to $\mathcal{O}(\epsilon^{7})$
when $\ell^{5}=576$). The continuation approach also allows the computation
of secondary branches of mixed modes (M) and folds (F).

\begin{figure}
\centering{}\includegraphics[width=0.75\linewidth]{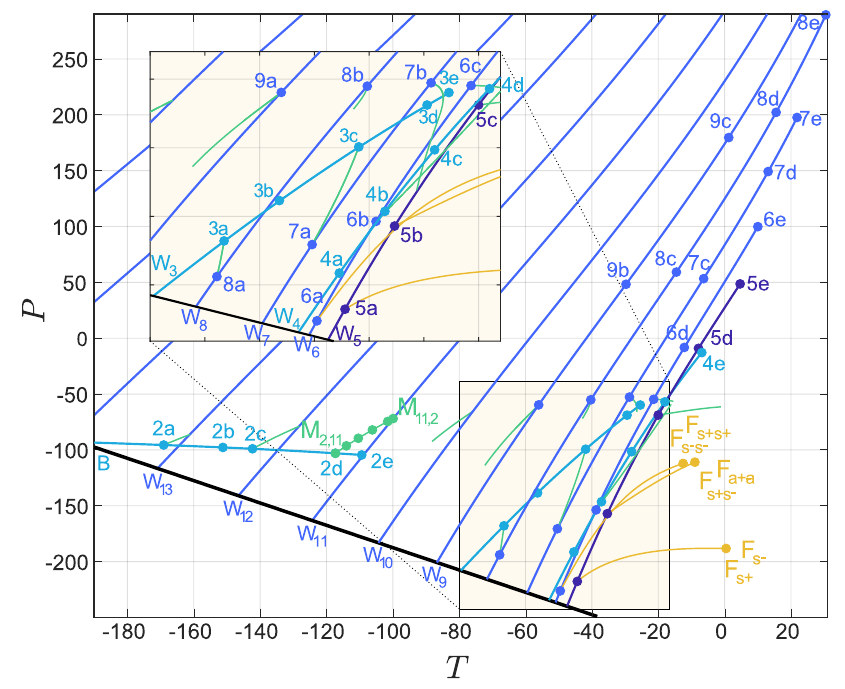}
\caption{Bifurcation diagram for $\ell^{5}=576$ (corresponding to $m^{*}=5$) showing the tension $T$
  resulting from an imposed pressure difference $P$ following the same colour scheme and labels as in
  figure~\ref{fig:PvsDelta}. The thick black line corresponds to the circle solution (\ref{eq:PT0}).
  The primary branch W$_5$ is shown as a thin purple line. The inset shows a zoom of the region near
  the primary bifurcation. The profiles corresponding to the labelled locations along each branch can
  be found in figure \ref{fig:PvsDelta}. An animation of the solutions along many of the solution
  branches in this figure is provided in the Supplementary Material \cite{supplementarymaterial}.
\label{fig:PvsT}}
\end{figure}

Figure~\ref{fig:PvsDelta} shows the compression $\Delta$ as a function of the
imposed pressure difference $P$ for primary wrinkle states W$_m$ with different
wavenumbers $m$, starting with W$_5$ corresponding to the onset wavenumber $m^*=5$.
The figure shows not only the pressure required to initiate collapse of the tube
(corresponding to $\Delta=0$) but also its subsequent response to quasistatic increase
in $P$, i.e., the figure represents the {\it tube law} describing the mechanical response
for different modes of instability for the chosen value $\ell^{5}=576$. Figure~\ref{fig:PvsT}
shows another measure of the response of the system, the tension $T$, also as a function of $P$.
The $(P,T)$ formulation provides the natural framework for numerical continuation.
Both figures also show a number of secondary branches (the mixed states M and the fold
states F) that bifurcate from the W states at finite amplitude, together with sample
solution profiles at the locations indicated in the figures. All our plots use the
same convention (colours and symbols).

While the circle solution ($\Delta=0$ in figure~\ref{fig:PvsDelta},
black line in figure~\ref{fig:PvsT}) exists for any pressure $P$,
we observe primary branches $\mathrm{W}{}_{m}$ of wrinkle states
with different integer wavenumbers $m$ only above the critical pressure
$P_{0}^{*}$. Wrinkle solutions with wavenumbers below $m^{*}$ are
interspersed with those above $m^{*}$; the wavenumber of the former
decreases as $P$ increases until $m=2$; thereafter only wrinkle
solutions with wavenumbers above $m^{*}$ are present and $m$ increases
monotonically with the pressure $P$. When $m^{*}$ is not an integer,
the primary instability corresponds to the integer $m$ nearest to
$m^{*}$ provided $m^*\ge 2$. Figure~\ref{fig:PvsDelta} shows that the compression $\Delta$
is almost proportional to the applied pressure $P$ for all the wrinkle
modes, i.e., that the modulus $Y\equiv\partial P/\partial\Delta$ is approximately
constant. Each $\mathrm{W}{}_{m}$ branch ultimately results in self-contact and at
this point the continuation is terminated. Self-contact forces can be included as
in \cite{flaherty_1972,flaherty_1973}, see also \cite{hazel_mullin_2017,pocheau_roman_2004},
but this has not been done here.

Besides wrinkle modes, numerical continuation reveals two types of
secondary branches. Most commonly, secondary branches connect a primary
mode with $m\geq m^{*}$ to another primary mode with $m<m^{*}$.
Figure~\ref{fig:PvsDelta} shows that all intermediate solutions along
the mixed-mode branch connecting $m=11$ and $m=2$ primary branches, i.e.
connecting the points $M_{11,2}$ to $M_{2,11}$, exhibit modulation at both
wavenumbers. In fact, most of these interconnecting branches also result
in self-contact, although longer, fully realisable interconnecting branches
become possible as $\ell$ (and hence $m^{*}$) increases and the number of
connections between W branches above and below $m^{*}$ grows.

Secondary bifurcations that do not connect different primary modes
are also present. These correspond to localised folds and come in
pairs. The first pair $\mathrm{F}{}_{s^{\pm}}$ bifurcates from $\mathrm{W}{}_{5}$
with $\mathrm{F}{}_{s^{+}}$ representing a localised protrusion while
$\mathrm{F}{}_{s^{-}}$ represents localised invagination. Both branches
reach self-contact at almost the same point (figures~\ref{fig:PvsDelta}
and \ref{fig:PvsT}). A family $\mathrm{F}_{a}$ of asymmetric
folds is also expected, but these states cannot be computed by AUTO
with the imposed boundary conditions. Arrays of folds with different
symmetries, analogous to those of \cite{LeoEdgarsheet}, have also
been found, with consistently higher degeneracy (see the yellow branches,
e.g. $\mathrm{F}{}_{s^{+}s^{-}}$ in figures~\ref{fig:PvsDelta} and
\ref{fig:PvsT}). Figure~\ref{fig:PvsDelta} also reveals that the
modulus $Y$ drops dramatically along the F branches, a well-known
consequence of the appearance of folds. In the case of the M branches,
the modulus $Y$ can be negative as is the case for the buckling mode
B.

\begin{figure}
\centering{}\includegraphics[width=0.75\columnwidth]{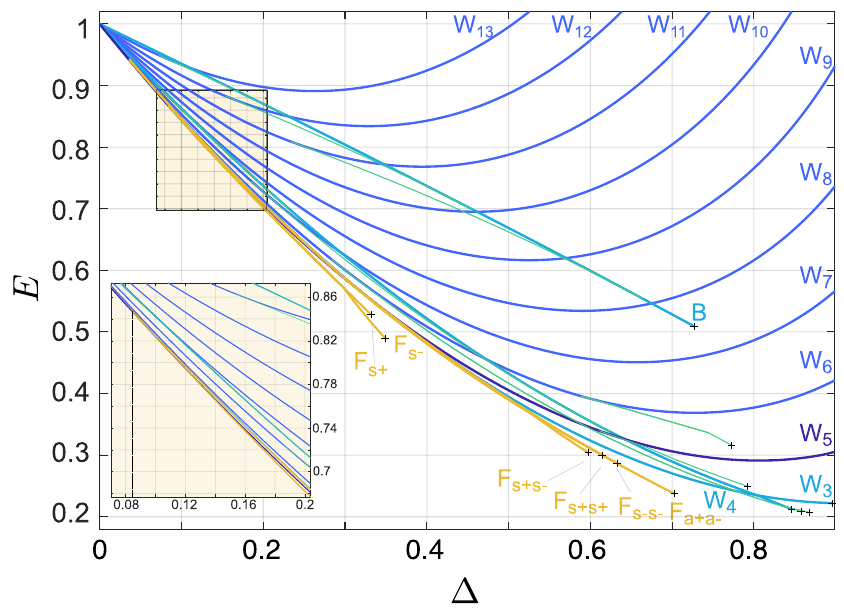}
\caption{The energy $E$ of the solutions in figures~\ref{fig:PvsDelta} and
\ref{fig:PvsT} across the full range of physical compression $\Delta$. Points
of self-contact are marked with crosses. All branches start from
the circle solution at $\Delta=0$. The inset shows a zoom of the region
where the fold state $\mathrm{F}{}_{s^{+}}$ becomes the global energy minimum.
\label{fig:EvsDelta}}
\end{figure}

We also examined the energy $E$ of the different wrinkled, folded
and buckled states as a function of the compression $\Delta$. For
small compression the lowest energy solution corresponds to $m^{*}=5$,
the natural wavenumber of the system for $\ell^5=576$, as shown in figure~\ref{fig:EvsDelta}.
However, as the compression increases, the localised states $\mathrm{F}{}_{s^{\pm}}$
bifurcate from the $m^{*}=5$ state, and the lowest energy state becomes
$\mathrm{F}{}_{s^{+}}$, with $\mathrm{F}{}_{s^{-}}$ at a slightly
higher energy. This secondary bifurcation thus defines the wrinkle-to-fold
transition, with threshold at $\Delta_{c}\approx0.084$ for the particular
case $\ell^{5}=576$. The direction of branching of $\mathrm{F}{}_{s^{+}}$
and $\mathrm{F}{}_{s^{-}}$ is consistent with that leading to spatially
localised states in the bistable Swift-Hohenberg equation \cite{burke2006}.
For higher compressions, the $\mathrm{F}{}_{s^{\pm}}$ are no longer
realisable and other localised states correspond to global energy minima
(figure~\ref{fig:EvsDelta}).

\begin{figure}
\centering{}\includegraphics[width=0.75\columnwidth]{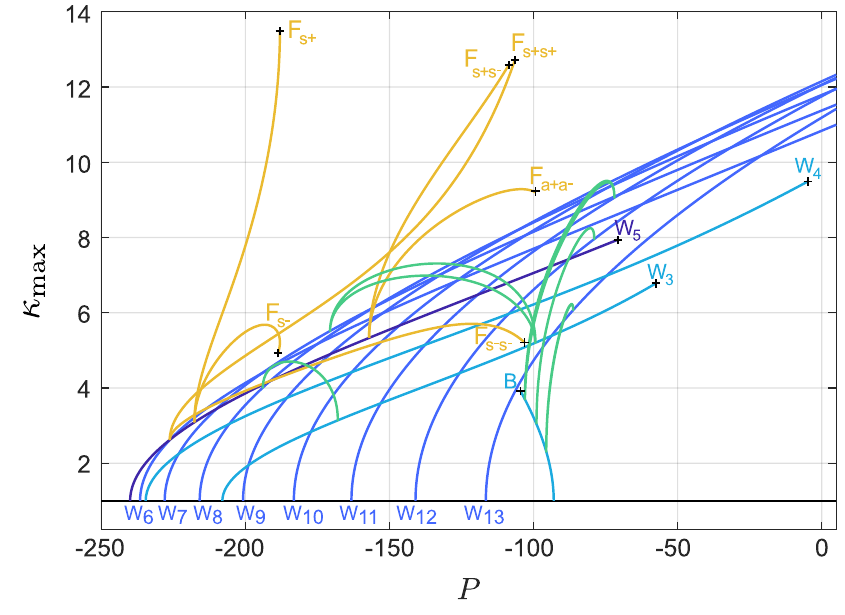}
\caption{Maximum curvature $\kappa_{\max}$ as a function of the pressure $P$
across the full range of compression for $\ell^{5}=576$. Self-contact
of the solutions is marked with crosses. The black horizontal line shows the
$R=1$ circle solution with $\kappa_{max}\equiv 1$. \label{fig:KappavsP}}
\end{figure}

Finally, in figure~\ref{fig:KappavsP} we plot the maximum curvature
of the different states we have studied. A rapid increase in maximum
curvature can be observed along all wrinkle branches after their bifurcation
from the constant curvature circle solution. Larger $m$ values result
in faster increase in $\kappa_{\max}$. Folds and some mixed states
display even faster increase in curvature after they emerge from secondary bifurcations.
The transition between the wrinkle state $\text{W}_{5}$ and the fold
states $\text{F}_{s^{\pm}}$, the first one to take place, occurs
at $P=-217.7$ (figure~\ref{fig:PvsT}) and corresponds to $\kappa_{\max}\approx3.09$.

\section{Numerical continuation: $\ell^5=0.005$}

When $\ell=0$ our problem becomes a pure buckling problem with no intrinsic length scale \cite{flaherty_1972,flaherty_1973}. In this case it is known that the first buckling mode corresponds to $m=2$ with more complex buckling modes requiring larger and larger pressures as the wavenumber $m$ increases. Moreover, in this regime the governing equation involves the curvature $\kappa$ only and the problem is analytically solvable \cite{vassilev_cylindrical_2008,Arreaga_shapes,guckenberger_theory_2017}.

\begin{figure}
\centering{}\includegraphics[width=0.99\linewidth]{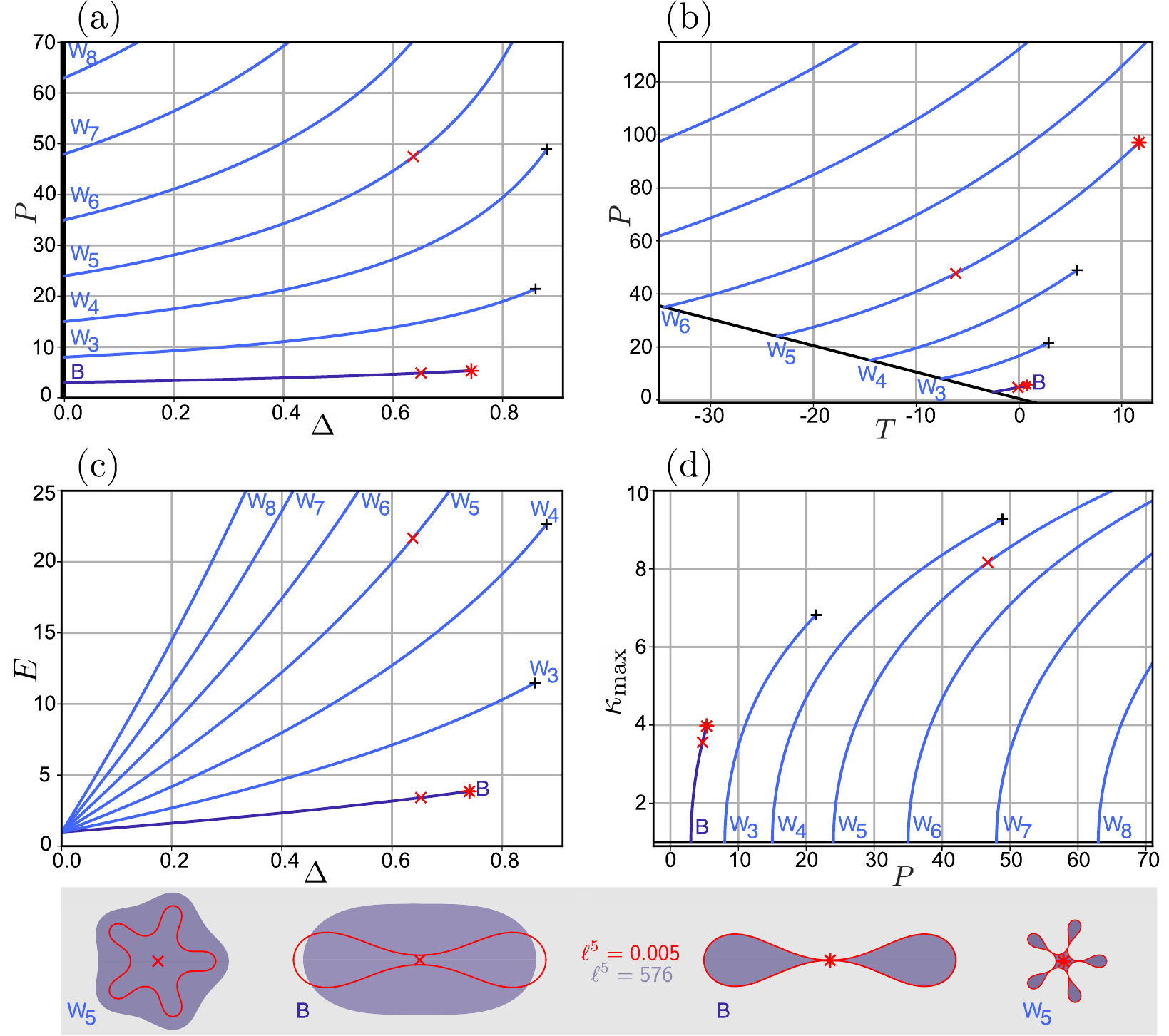}
\caption{Bifurcation diagrams for $\ell^{5}=0.005$ for comparison with the case $\ell^5=576$ (figures \ref{fig:PvsDelta}--\ref{fig:KappavsP}) showing the circle state in black and all bifurcations from it, following the same colour scheme and labels as in figure~\ref{fig:PvsDelta}. The first primary branch is now the $m=2$ buckling mode (B, thin purple line). Solution branches are shown up to the point of self-contact (crosses).
(a) The compression $\Delta$ as a function of the pressure $P$. The modulus $Y=\partial P/\partial\Delta$ is much smaller than in figure~\ref{fig:PvsDelta}, but positive for all wavenumbers $m$ including $m=2$.
(b) The tension $T$ resulting from an imposed pressure difference $P$. All primary branches have positive slope $\partial P/\partial T$ in accord with the weakly nonlinear theory. 
(c) The energy $E$ as a function of the compression $\Delta$. The energy increases with increasing $\Delta$ for all primary branches until self-contact; the $m=2$ buckled state (B, thin purple line) is the global minimum energy state.
(d) The maximum curvature $\kappa_{\rm max}$ as a function of $P$. The circle solution corresponds to $\kappa_{\rm max}=1$ (black horizontal line). Self-contact is reached for much smaller pressure changes than for $\ell^5=576$.
  The bottom two panels depict the $m=2$ and $m=5$ solution profiles when $P$ is increased from $P_0$ by 10\% ($\Delta P=24$, $\ell^5=576$) and 2\% ($|\Delta P|=1.86$, $\ell^5=0.005$). This point is indicated by a red $\times$ in the figure: the tube is substantially more compressed when $\ell^5$ is smaller.  We also depict the $m=2$ and $m=5$ solutions overlaid at the point of self-contact for both $\ell$ values (red $*$, the only point at which different $\ell$ values can be precisely compared). The figure shows that for both $m=2$ and $m=5$ the profiles at this point are identical, i.e., the profiles at the point of contact are independent of $\ell$.
\label{fig:l5005}}
\end{figure}

To confirm that our model possesses the correct limiting behaviour and thereby validate our numerical continuation approach we take $\ell^5$ to be very small and compare our results with those for $\ell=0$ and $\ell^5=576$. Specifically, we take $\ell^5=0.005$ and document the corresponding nonlinear results in Figure \ref{fig:l5005} for comparison with figures \ref{fig:PvsDelta}--\ref{fig:KappavsP}.

As expected, the first solution to emerge from the circle when $\ell^5=0.005$ is $m=2$, i.e., the buckling mode B, and the wavenumber of the subsequent solutions that emerge increases monotonically with the pressure difference $P$. Moreover, the appearance of these states requires positive values of $P$ and the corresponding branches all behave in a similar fashion. These solutions are thus the expected buckled states. In figure \ref{fig:l5005} these states are still labelled W but this is only because $\ell$ is not identically zero. For these small values of $\ell$ there are no mixed modes or folds prior to self-contact. In fact, such secondary bifurcations move farther and farther out along each primary branch and beyond the point of self-contact as $\ell^5\to0$, and conversely, down each primary branch and towards the circle solution when $\ell^5$ increases. This process leads, for sufficiently large $\ell$, to the appearance of secondary bifurcations prior to self-contact and for negative values of $P$, as in figure \ref{fig:PvsT}. 

Figure~\ref{fig:l5005}a shows the compression $\Delta$ as a function of $P$ for $\ell^5=0.005$ for comparison with figure~\ref{fig:PvsDelta} while figure~\ref{fig:l5005}b shows the tension $T$, also as a function of $P$. Figure \ref{fig:l5005}a reveals that for smaller values of $\ell^5$ the compression increases much more rapidly with $P$ than for larger values $\ell^5$, a consequence of the absence of the stiffening effect of the substrate. These results are corroborated in \ref{fig:l5005}b. The results in both figures are in accord with the weakly nonlinear theory: the modulus $Y=\partial P/\partial\Delta$ is now positive for all wavenumbers $m$ (figure \ref{fig:l5005}a) and likewise all primary branches have positive slopes $\partial P/\partial T$ (figure \ref{fig:l5005}b), even for $m=2$, as predicted by the theory. 

Figure~\ref{fig:l5005}c shows the energy $E$ as a function of the compression $\Delta$ for $\ell^{5}=0.005$ for comparison with figure~\ref{fig:EvsDelta}. In contrast to the latter, $E$ is now a monotonically increasing function of $\Delta$ and the wavenumber $m$: for small $\ell^5$ the bending energy dominates the substrate energy and its contribution grows with increasing compression. Thus the lowest energy state at a given compression is that with the lowest overall curvature, i.e., the wavenumber $m=2$ state is the minimum energy state and so is stable until self-contact (cross). After this point, stability is transferred to the next lowest wavenumber solution, $m=3$, etc.

In figure~\ref{fig:l5005}d, we plot the maximum curvature as a function of $P$ for comparison with figure~\ref{fig:KappavsP}. The figure shows that for small $\ell^5$ maximum curvature is reached much earlier as $P$ increases than for larger $\ell^5$. However, in each case, the maximum value of $\kappa_{\rm max}$ necessarily coincides with the point of self-contact and is identical to the corresponding curvature when $\ell=576$, i.e., $\kappa_{\rm max}$ is independent of $\ell$ (figure 7, bottom right panel).

All this is in substantial contrast to the behaviour identified at larger $\ell^5$ described in figures \ref{fig:PvsDelta}--\ref{fig:KappavsP} but confirms that the solutions of (\ref{eq:main}) converge to the correct pure buckling limit as $\ell\to0$.

Finally, the two lowest panels in figure~\ref{fig:l5005} compare the profiles of the $m=2$ and $m=5$ solutions for the two different values of $\ell$ considered in this work. The comparison is made at a point 10\% from the critical pressure $P_0$ for $m=5$ and 2\% from the critical pressure for $m=2$ and again at the point of self-contact for both (red $\times$ and $*$ symbols, respectively). We see that when $\ell$ is large the amount of compression for given $\Delta P$ is substantially less than for smaller $\ell$. Thus the wrinkling or buckling process occurs over a smaller interval of $P$ as $\ell$ decreases. However, at the point of self-contact the profiles in the two cases are identical and independent of the parameter $\ell$ as suggested by the weakly nonlinear analysis. 

Evidently, as $\ell^5$ decreases and the influence of the substrate wanes the bifurcation diagrams simplify dramatically and in the absence of the second length scale the system approaches the corresponding result for the unsupported ring ($\ell=0$). This simplification arises because the secondary branches leading to both mixed modes and the folded states move past the point of self-contact thereby ceasing to be realisable. In this case the first primary mode is the lowest wavenumber mode, $m=2$. Subsequent primary modes now come in monotonically with increasing $m$ and all behave in a similar fashion. However, despite these changes the primary branches continue to bifurcate subcritically, in the sense that the lining loosens (tension $T$ becomes less negative), as $P$ increases.

\begin{figure}
  \centering{}\includegraphics[width=0.99\linewidth]{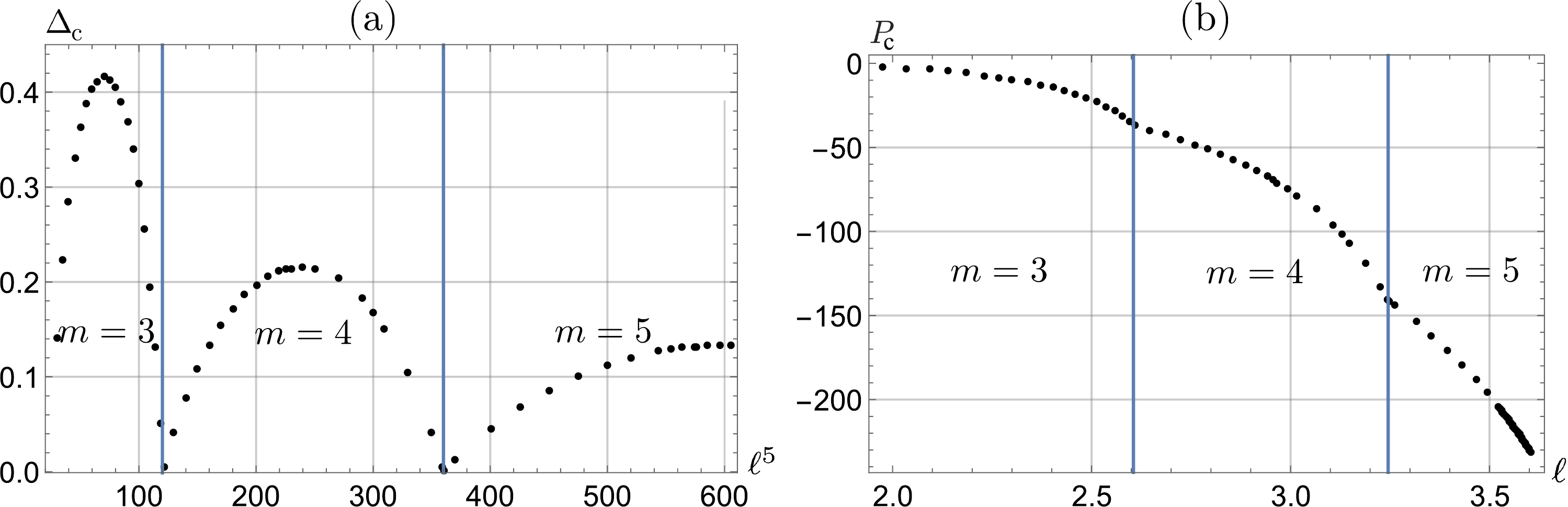}
\caption{(a) The compression $\Delta_c$ at the secondary bifurcation from the primary wrinkled state to the first fold state as a function of $\ell^5$, showing the behaviour of this bifurcation point with changing wavenumber of the wrinkled state. The vertical lines show the codimension-two points $\ell^5_{m,m+1}\equiv (m-1)m(m+1)(m+2)$ when $m=3$ and 4. (b) The corresponding plot of $P_c$ as a function of $\ell$.
  \label{fig:cd2}}
\end{figure}

On the other hand when $\ell^5$ increases the wavenumber $m^*$ of the mode that first sets in also increases (figure \ref{fig:P0}). This fact leads to repeated {\it mode jumping}. For example, $m^*=4$ for $\ell^5=320$ while $m^*=5$ for $\ell^5=576$. Thus the mode $m^*=4$ remains dominant only over a finite interval of $\ell^5$, and as $\ell^5$ increases $m^*=4$ is replaced by a new dominant mode, $m^*=5$. This transition is associated with a so-called codimension-two point where the dispersion relation (\ref{eq:quintic_m}) is simultaneously solved by two adjacent values of $m$, here $m^*=4$ and $m^*=5$. A similar situation occurs in the planar case, as described in detail in \cite[Figure 5]{LeoEdgarsheet}. In particular, when $m^*=4$ the folds F bifurcate from W$_4$; as $\ell^5$ increases towards the codimension-two point $\ell^{5}_{4,5}=360$ the secondary bifurcations leading to the folds move down along the W$_4$ branch and reach zero amplitude when $\ell^5=\ell^5_{4,5}$. For $\ell^5>\ell^5_{4,5}$ the dominant mode is $m^*=5$ and the secondary bifurcation to the fold state now takes place on W$_5$. As $\ell^5$ increases this bifurcation moves up along W$_5$ to a maximum amplitude before moving down again as the next codimension-two point is approached. This process repeats as $\ell^5$ continues to increase, and $\Delta_c$, the threshold for the onset of the fold state, therefore both oscillates and jumps from branch to branch. This behaviour is shown in figure \ref{fig:cd2}(a) and is similar to that found in the planar case \cite[Figure 5]{LeoEdgarsheet}; we expect that the tube problem studied here approaches the planar case once $\ell^5$ is sufficiently large (sufficiently large tube radius).

\section{Conclusion}

In this article, we provided a simple model of an inextensible elastic
lining of an inner-lined tube subjected to an imposed pressure difference,
and described its buckled, wrinkled and folded solutions. We showed
that wrinkling is statically generated by a competition between bending,
soft-substrate forces and the applied pressure, and explored the limiting
behaviour of our model as the strength of the substrate support is reduced
eliminating the possibility of wrinkling. We showed that for sufficiently
strong substrate support, increasing the applied pressure leads not only
to a wrinkle-to-fold transition, but also to mixed states. The energies of
these states were calculated using weakly nonlinear theory and by numerical
continuation for strongly nonlinear solutions. The wrinkle state with wavelength
closest to natural is initially the state with the least energy and is thus stable
until a single-fold state bifurcates from it.  As $\ell^5$ increases, additional
mixed modes arise prior to self-contact, and states with an increasing number of
localised folds become possible. The solution profiles match well with observations
and resemble structures in growing composite rings \cite{michaels_puckering_2021}.
Our approach explains how the wavelength and amplitude of the wrinkles are selected
as a function of parameters in pressure-driven wrinkling systems. This is in turn key to
understanding, for example, the artery self-cleaning process arising from wrinkled-to-unwrinkled
cycles triggered by blood pressure changes \cite{pocivavsek2018,pocivavsek2019}
and can be a good starting point for more refined models that include adhesion.
A natural question that arises is how the bending modulus, the size
of the system and the substrate properties may be optimised to maximise
in-plane curvature, thereby optimising the self-cleaning properties
for a given pressure jump, while avoiding the wrinkle-to-fold transition.
For weaker substrate support the first primary bifurcation is to the $m=2$
buckling mode, and the secondary bifurcations move to large amplitudes,
beyond the point of self-contact. Thus all bifurcation diagrams simplify
and wavenumber of the primary branches increases monotonically with increasing
pressure.

Applications of this work to the time-dependent artery problem and to other systems
exhibiting competition between buckling, wrinkling and folding will be described elsewhere.

\ack{}{This work was supported in part by the National Science Foundation
under grant DMS-1908891 (BF, NV \& EK). The work of NV was funded by the
National Agency for Research and Development (ANID) through the Scholarship
Program: Becas de Postdoctorado en el Extranjero, Becas Chile 2018
No. 74190030. LG was funded by grant Conicyt Fondecyt Iniciaci\'{o}n 11170700.
We thank E. Cerda for valuable discussions. }

\appendix

\section{Derivation from Kirchhoff equations\label{sec:AppendixA}}

In equilibrium, the forces acting on an element of the lining can
be expressed in terms of the static Kirchhoff equations \cite{kodio_weak_nonlin,audoly_pomeau}: 
\begin{eqnarray}
\partial_{s}\mathbf{F}+\left[P-\frac{1}{2}K\left(r_{0}^{2}-r^{2}\right)\right]\mathbf{n} & = & 0,\label{eq:Kirchoff1}\\
\qquad\qquad\qquad\partial_{s}\mathbf{M}+\mathbf{t}\times\mathbf{F} & = & 0,\label{eq:Kirchoff2}
\end{eqnarray}
where $\mathbf{F}$ and $\mathbf{M}$ are the force and moment acting
on the centerline of the element. The extra pressure in square brackets
in (\ref{eq:Kirchoff1}) is due to the force per unit of area exerted
by the substrate, modelled by a Winkler foundation \cite{winkler} with a nonlinear
quadratic term.

The moment is related to the local curvature of the ring by the constitutive
relation $\mathbf{M}=\left(\mathcal{B}\partial_{s}\phi\right)\mathbf{k}$,
where $\phi$ is the angle between a tangent to the ring and fixed
horizontal axis, and $\mathbf{k}$ is normal to the plane. Thus $\partial_{s}\mathbf{r}=\left(\cos\phi,\sin\phi\right)$.
The unit vectors $\mathbf{t}$ and $\mathbf{n}$ are given by $\mathbf{t}=\partial_{s}\mathbf{r}=\left(\cos\phi,\sin\phi\right)$
and $\mathbf{n}=\left(-\sin\phi,\cos\phi\right)$; \textbf{$\mathbf{n}$}
points towards the interior of the enclosed region. Accordingly, the
system can be written in terms of five differential equations: 
\begin{eqnarray*}
\quad\partial_{s}x & = & \cos\phi,\\
\quad\partial_{s}y & = & \sin\phi,\\
\mathcal{B}\partial_{ss}\phi & = & F_{x}\sin\phi-F_{y}\cos\phi,\\
\quad\partial_{s}F_{x} & = & +\left[P-\frac{1}{2}K\left(r_{0}^{2}-r^{2}\right)\right]\sin\phi,\\
\quad\partial_{s}F_{y} & = & -\left[P-\frac{1}{2}K\left(r_{0}^{2}-r^{2}\right)\right]\cos\phi.
\end{eqnarray*}
The problem is defined after imposing the closed-curve boundary condition
$\phi\left(L,t\right)=\phi\left(0,t\right)+2\pi$, $L=2\pi R$, and
periodic boundary conditions on $\partial_{s}\phi,x,y,F_{x}$ and
$F_{y}$. We show that this system of equations is equivalent to (\ref{eq:phiODE})
of the text. For this purpose, we rewrite (\ref{eq:Kirchoff2}) using
the constitutive relation $\mathbf{M}=\left(\mathcal{B}\partial_{s}\phi\right)\mathbf{k}$
and the identity $\left(\partial_{s}\phi\right)\mathbf{k}=\partial_{s}\mathbf{r}\times\partial_{ss}\mathbf{r}$,
\[
\partial_{s}\mathbf{r}\times\left(B\partial_{sss}\mathbf{r}+\mathbf{F}\right)=0,
\]
which is solved by $\mathbf{F}=-\mathcal{B}\partial_{sss}\mathbf{r}+\lambda\partial_{s}\mathbf{r}$,
where we have introduced the Lagrange multiplier $\lambda(s)$ to
incorporate inextensibility. The latter expression, substituted in
(\ref{eq:Kirchoff1}), yields 
\[
-\mathcal{B}\partial_{sss}\mathbf{r}+\lambda\partial_{ss}\mathbf{r}+\partial_{s}\lambda\partial_{s}\mathbf{r}+\left[P-\frac{1}{2}K\left(r_{0}^{2}-r^{2}\right)\right]\mathbf{n}=0.
\]
To simplify this expression, we use the following identities: $\partial_{s}\mathbf{t}=\mathbf{\left(\partial_{s}\phi\right)\mathbf{n}}$,
$\partial_{s}\mathbf{n}=\mathbf{-\left(\partial_{s}\phi\right)\mathbf{t}},$
$\partial_{s}\mathbf{r}=\mathbf{t}$, $\partial_{ss}\mathbf{r}=\left(\partial_{s}\phi\right)\mathbf{n}$,
$\partial_{sss}\mathbf{r}=\left(\partial_{ss}\phi\right)\mathbf{n}-\left(\partial_{s}\phi\right)^{2}\mathbf{t}$
and $\partial_{ssss}\mathbf{r}=\left(\partial_{sss}\phi-\left[\partial_{s}\phi\right]^{3}\right)\mathbf{n}-\left(3\partial_{s}\phi\partial_{ss}\phi\right)\mathbf{t}$.
The result is 
\begin{eqnarray*}
\left(-\mathcal{B}\partial_{sss}\phi+\mathcal{B}\left[\partial_{s}\phi\right]^{3}+\lambda\partial_{s}\phi+P-\frac{1}{2}K\left(r_{0}^{2}-r^{2}\right)\right)\mathbf{n}+\ldots\\
\qquad\qquad\qquad\qquad\qquad\qquad\ldots\left(\frac{}{}3\partial_{s}\phi\partial_{ss}\phi+\partial_{s}\lambda\right)\mathbf{t} & = & 0.
\end{eqnarray*}
Since $\mathbf{n}$ and $\mathbf{t}$ form an orthonormal basis, the
two terms in parentheses must both vanish. From the second, we obtain
a differential equation for $\lambda$ whose solution is 
\[
\lambda\left(s\right)=-\frac{3}{2}\left(\partial_{s}\phi\right)^{2}+T,
\]
where $T$ is a constant. Replacing $\lambda\left(s\right)$ in the
first set of parentheses by the above expression, we finally obtain:
\begin{equation}
-\mathcal{B}\partial_{sss}\phi-\frac{1}{2}B\left(\partial_{s}\phi\right)^{3}+T\partial_{s}\phi+P-\frac{1}{2}K\left(r_{0}^{2}-r^{2}\right)=0,\label{eq:main-1}
\end{equation}
leading to eq.~(\ref{eq:phiODE}).

The same equation can also be derived from a constrained Lagrangian as done for the planar elastic sheet in \cite{diamant_witten2011}.

\section{Weakly nonlinear analysis\label{sec:AppendixB}}

At each order in the weakly nonlinear analysis, we obtain a linear problem of the form 
\[
\mathcal{L}[\phi_{j},x_{j},y_{j}]\equiv\partial_{s}^{3}\phi_{j}+\left(\frac{3}{2}-T_{0}\right)\partial_{s}\phi_{j}-\ell^{5}\left(x_{0}x_{j}+y_{0}y_{j}\right)=\mathcal{N}_{j},
\]
for $j=1,2,...$, with the first three $\mathcal{N}_{j}$ given by
\begin{eqnarray*}
\mathcal{N}_{1}= & 0\\
\mathcal{N}_{2}= & - & \left(\frac{3}{2}(\partial_{s}\phi_{1})^{2}+\frac{1}{2}\ell^{5}(x_{1}^{2}+y_{1}^{2})-P_{2}-T_{2}\right)\\
\mathcal{N}_{3}= & - & \left(\frac{1}{2}(\partial_{s}\phi_{1})^{3}+3(\partial_{s}\phi_{1})(\partial_{s}\phi_{2})+\ell^{5}(x_{1}x_{2}+y_{1}y_{2})\right)+T_{2}\partial_{s}\phi_{1}.
\end{eqnarray*}
To eliminate $x_{j}$ and $y_{j}$ from $\mathcal{L}[\phi_{j},x_{j},y_{j}]$,
we compute $(\partial_{s}^{2}\mathcal{L}+\mathcal{L})[\phi_{j},x_{j},y_{j}]$:

\begin{eqnarray*}
\fl\partial_{s}^{5}\phi_{j}+\left(\frac{5}{2}-T_{0}\right)\partial_{s}^{3}\phi_{j}+\left(\frac{3}{2}-T_{0}\right)\partial_{s}\phi_{j}+\ldots & \ \\
\ldots\ell^{5}\left[2(\partial_{s}x_{0})(\partial_{s}x_{j})+2(\partial_{s}y_{0})(\partial_{s}y_{j})+x_{0}\partial_{s}^{2}x_{j}+y_{0}\partial_{s}^{2}y_{j}\right] & = & (\partial_{s}^{2}+1)\mathcal{N}_{j}.
\end{eqnarray*}
Expansion of the geometric identities $\partial_{s}x=\cos\phi$ and $\partial_{s}y=\sin\phi$ now results in 
\begin{equation}
\partial_{s}^{5}\phi_{j}+\left(\frac{5}{2}-T_{0}\right)\partial_{s}^{3}\phi_{j}+\left(\frac{3}{2}-T_{0}+\ell^{5}\right)\partial_{s}\phi_{j}=\mathcal{G}_{j}+(\partial_{s}^{2}+1)\mathcal{N}_{j},\label{eq:wnla}
\end{equation}
where the first three $\mathcal{G}_{j}$ are given by 
\begin{eqnarray*}
 & \mathcal{G}_{1} & =0\\
 & \mathcal{G}_{2} & =\frac{1}{2}\ell^{5}(\partial_{s}\phi_{1})^{2}\\
 & \mathcal{G}_{3} & =\ell^{5}\left(\frac{1}{2}\phi_{1}^{2}\partial_{s}\phi_{1}-\phi_{1}\phi_{2}\right).\\
\end{eqnarray*}
Solving (\ref{eq:wnla}) for $j=1,2$ subject to the requirement that the solution is periodic
yields the expressions for $\phi_{1},A_{1}$ and for $\phi_{2},A_{2},P_{2},T_{2}$ given in the
text. For $j$ even, the solvability condition
imposed on $\mathcal{G}_{j}+(\partial_{s}^{2}+1)\mathcal{N}_{j}$
generates $P_{j}(T_{j})$, while for $j$ odd, it generates $T_{j-1}(m,\ell^{5})$.
Higher order expressions were obtained through symbolic calculations
using the software Maple.
\begin{figure*}
\centering{}\includegraphics[width=0.5\paperwidth]{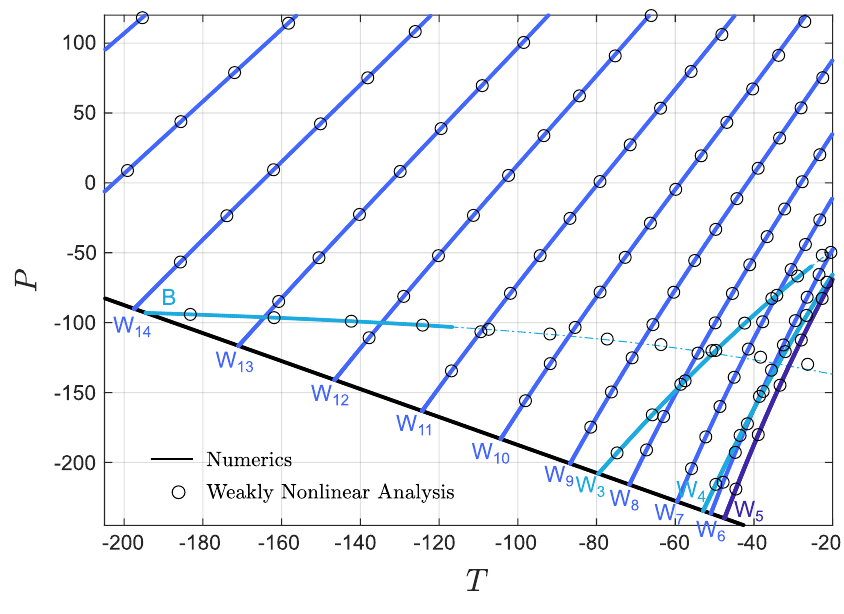}
\caption{Comparison between numerical continuation for $\ell^5=576$ (solid lines) and the corresponding
  $\mathcal{O}(\epsilon^{7})$ weakly nonlinear analysis (open circles) demonstrating excellent agreement
  between perturbation theory and numerically exact solutions extending to $\epsilon=\mathcal{O}(1)$ at
  the top of the figure (cf.~figure \ref{fig:BD_WNLA_sol} and \ref{sec:AppendixD}). \label{fig:BD_WNLA}}
\end{figure*}

\section{Numerical continuation with AUTO\label{sec:AppendixC}}

We implemented the problem (\ref{eq:main}) in AUTO \cite{doedel08auto-07p}
as a 5-dimensional boundary value problem on the domain $s\in[0,\pi]$,
representing one half of the lining, with the boundary conditions
$\phi\left(0\right)=\pi/2$, $\phi\left(\pi\right)=3\pi/2$, $x\left(0\right)=x_{0}$,
$x\left(\pi\right)=x_{1}$ and $y\left(0\right)=y\left(\pi\right)=0$
together with the force-free conditions $\phi''\left(0\right)=\phi''(\pi)=0$
\cite{kodio_weak_nonlin}. The boundary conditions constrain the rotation
symmetry in $\phi$ and eliminate translations in $y$, while the
force-free boundary conditions permit reflection in $y=0$ to generate
solutions on the full circle. A 5-dimensional system with 8 boundary
conditions requires 4 degrees of freedom in the parameters \cite{Doedel_1991},
so we perform our continuation in $\left(P,T,x_{0},x_{1}\right)$.
This procedure allows $T$ to adjust to increments in $P$ and the
endpoints $x_{1}$, $x_{2}$ to change in accordance with the zero-force
condition. Figure 3 of the text shows the resulting full circle profiles.

The imposed boundary conditions prevent the computation
of asymmetric states F$_{a}$ that are also expected to appear via secondary
bifurcations from wrinkled states. 

\begin{figure*}
\centering{}\includegraphics[width=0.5\paperwidth]{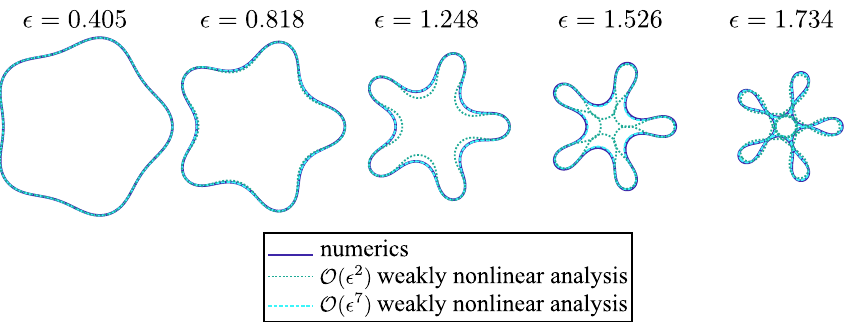}
\caption{Comparison between numerical solutions on the W$_5$ branch and weakly nonlinear solutions
at $\mathcal{O}(\epsilon^{2})$ and $\mathcal{O}(\epsilon^{7})$ when $\ell^5=576$.\label{fig:BD_WNLA_sol}}
\end{figure*}

\section{Comparisons\label{sec:AppendixD}}

In figure~\ref{fig:BD_WNLA}, we compare the results from numerical
continuation of (\ref{eq:phiODE}) of the text with the above boundary
conditions and the corresponding results obtained above from weakly
nonlinear theory carried out to $\mathcal{O}(\epsilon^{7})$. The
$\epsilon$ values corresponding to the maximum displayed extent of
each branch are summarized in table~\ref{tab:table1}. The results
demonstrate excellent agreement between perturbation theory and the
numerically exact solutions for $\epsilon\lesssim1$. Equally good
agreement is found for the solution profiles as shown in figure~\ref{fig:BD_WNLA_sol}.

\begin{table}[h]
\caption{Wavenumber $m$ and maximum $\epsilon$ used in figure~\ref{fig:BD_WNLA}.\label{tab:table1} }

\centering{}%
\begin{tabular}{|>{\centering}p{0.5cm}|>{\centering}p{0.5cm}|>{\centering}p{0.5cm}|>{\centering}p{0.5cm}|>{\centering}p{0.5cm}|>{\centering}p{0.5cm}|>{\centering}p{0.5cm}|>{\centering}p{0.5cm}|>{\centering}p{0.5cm}|>{\centering}p{0.5cm}|>{\centering}p{0.5cm}|>{\centering}p{0.5cm}|>{\centering}p{0.5cm}|>{\centering}p{0.5cm}|}
\hline 
{\small{}m} & {\small{}2} & {\small{}3} & {\small{}4} & {\small{}5} & {\small{}6} & {\small{}7} & {\small{}8} & {\small{}9} & {\small{}10} & {\small{}11} & {\small{}12} & {\small{}13} & {\small{}14}\tabularnewline
\hline 
{\small{}$\epsilon_{\textrm{max}}$} & {\small{}1.96} & {\small{}1.54} & {\small{}1.32} & {\small{}1.17} & {\small{}1.25} & {\small{}1.29} & {\small{}1.35} & {\small{}1.37} & {\small{}1.37} & {\small{}1.25} & {\small{}1.18} & {\small{}1.04} & {\small{}0.96}\tabularnewline
\hline 
\end{tabular}
\end{table}

\vspace{0.4cm}
 \section*{References}{\bibliographystyle{unsrt}
\bibliography{main}
}
\end{document}